\documentstyle[12pt]{article}
\renewcommand{\slash}[1]{/\kern-7pt #1}

\def\N1{N=1 supersymmetric }

\def\HV{`t~Hooft-Veltman }

\def\DR{dimensional reduction }

\def\1PI{one-particle irreducible }

\def\tr{{\rm Tr}}
\def\d{{\rm d}}
\def\e{{\rm e}}
\def\i{{\rm i}}
\def\A{{\cal A}}
\def\F{{\cal F}}
\def\I{{\cal I}}
\def\M{{\cal M}}
\def\S{{\cal S}}
\def\ds{\displaystyle}
\def\to{\rightarrow}
\def\ms{$\overline{{\rm MS}}$}
\def\lQCD{$\Lambda_{{\rm QCD}}$}
\def\l{\langle}
\def\r{\rangle}
\def\vspaceinarray{\nonumber ~&~&~\\}

\begin{document}

\begin{titlepage}
\vspace*{-2cm}
\begin{flushright}
ETH-TH/94-3\\
January 1994 \\
\end{flushright}
\vskip .5in
\begin{center}
{\Large\bf
Singular terms of   helicity amplitudes at one-loop in QCD
and the soft limit of the cross sections of multi-parton
 processes}\footnote{
Work supported in part by the Schweizerischer
Nationalfonds}\\
\vskip 1cm
{\large Zoltan Kunszt, Adrian Signer and Zolt\'an  Tr\'ocs\'anyi} \\
\vskip 0.2cm
Theoretical Physics, ETH, \\
Z\"urich, Switzerland  \\
\vskip 1cm
\end{center}

\begin{abstract}
\noindent
We describe a general  method that enables us to obtain
all the singular terms of  helicity amplitudes of n-parton
processes at one loop. The  algorithm uses  helicity amplitudes
at tree level and
simple color algebra.
 We illustrate the method
by calculating the
singular part of the one loop helicity amplitudes of all
$ 2\to 3$ parton subprocesses.
The results  are used to derive the soft gluon limit of
the cross sections of all $2\to 4$ parton scattering subprocesses
which provide a useful initial condition for the angular ordering
approximation to coherent  multiple soft gluon emission,
incorporated in existing Monte Carlo simulation programs.
\end{abstract}
\end{titlepage}
\setcounter{footnote}{0}

\bigskip

\section{Introduction}
The technical developments achieved recently in perturbative QCD calculations
(helicity method
\cite{Xu87,Gun85,KleissWJS85,Man91},
string theory based
derivations \cite{Ber92,Ber93}) enable us to
calculate one-loop  corrections to helicity amplitudes
up to five-parton  and perhaps also to six-parton processes.
Some partial  results for the five parton amplitudes
 are by now  available.

In this development the utilization of  the  helicity method was
decisively important.
 It was used both in  the    string theory based derivation of the one-loop
corrections to the
four-gluon \cite{Ber92} and five-gluon \cite{Ber93} helicity amplitudes,
as well as in
the recent  conventional diagrammatic calculation of the  one
loop corrections of  the  helicity amplitudes for
all $2\to 2$ parton processes\cite{ZKASZT93}.

The helicity method which
 was originally proposed for
calculating tree amplitudes \cite{Man91}
 can be extended
 to the calculation of loop corrections by observing
that the difficulties given by the appearance of $\gamma_5$ in the
definition of the helicity states can be circumvented if instead of
conventional dimensional regularization
  the \HV regularization \cite{'tH72,Gas73} or
  dimensional reduction
\cite{Sie79,Cap80,Sie80,Alt81}
 are utilized.
While employing these regularizations requires some care
in  calculating   two- or more-loop contributions,
 their use is rather
straightforward
 in the  case of  one-loop corrections.
The regularization dependent terms
in the loop contributions  obtained with
conventional dimensional regularization, \DR or \HV regularization
are determined by  the  universal
process independent structure of the singular contributions. The
process independent transition rules
have been worked out recently \cite{ZKASZT93}.

An important technical  issue in the evaluation  of next-to-leading
 QCD corrections
to physical cross-section
is the analytic cancelation of     the soft and collinear
singularities between the loop  and the
Bremsstrahlung contributions.
 The singular terms  in the
cross-sections
coming from  Bremsstrahlung contributions have universal, process independent
structure.
 They depend only on
the color representation of the external legs
and each leg   contributes additively.
 It was recently shown that
 these terms
  can easily be obtained   with the help of a Born-level
algorithm \cite{Kun92} (see also \cite{Cia81}).
Since the virtual corrections are cancelled by the Bremsstrahlung
contributions,
one expects
that  a  simple Born-level algorithm  should also be sufficient
to calculate directly the singular terms of the virtual corrections as well.
 The aim of this paper
is to show that indeed
such an algorithm exists\footnote{ A comprehensive treatment of the soft
limit in QCD is presented in\cite{Cia81}. For a recent paper
in the large $N_c$ limit see \cite{Gie92}.}.

In section 2 we  give a brief  description of our algorithm for
calculating
the singular terms of  one-loop helicity amplitudes.
In section 3
 we illustrate the power of the method
by calculating
 the singular terms of  one-loop helicity amplitudes of
all    $2\to 3$ subprocesses.
  The soft singular terms  of
the one-loop helicity amplitudes of the $2\to n-2$
parton processes
determine also
 the soft limit of the cross sections
of the $2\to n-1$ parton scattering processes via the cancelation
theorem. In view of its phenomenological usefulness, we
summarize our results for  the soft limit  of the Born cross sections
 of all $2\to 4 $ subprocesses
in Section 4.


\section{Singular terms in one-loop helicity amplitudes}
\setcounter{equation}{0}
In $D$-dimensional regularisation the virtual corrections  to
helicity  amplitudes at one loop have soft
$1/\epsilon^2$ and $1/\epsilon$, collinear
$1/\epsilon $
and ultraviolet
$1/\epsilon$
 singularities. The ultraviolet singular term is trivially cancelled
by the \ms\  coupling constant counter term
 \begin{equation}
\label{UVsing}
{\cal A}^{\rm UV}=-{{(n-2)(4\pi)^{\epsilon}}
\over {2\epsilon\Gamma(1-\epsilon)}} \left(g\over 4\pi\right)^2
\beta_0 \  {\cal A}^{\rm tree}\,
\end{equation}
where $n$ denotes the number of the
external legs, $\beta_0$ is the first term in the $\beta$-function
 \begin{equation}
\beta_0=(11 N_c-2 N_f)/3\, ,
\end{equation}
  $g$ denotes the renormalized coupling constant and $ D = 4- 2
\epsilon$.  Color and
helicity labels are supressed.
In the following we shall discuss only renormalized
helicity amplitudes.

In axial gauge the collinear singularities come from the self
energy corrections to the external lines \cite{ColSopSt87}.
For each helicities    and color subamplitudes
  they are
 proportional to the Born term
since  the
Altarelli-Parisi  functions, $P_{a/a}(z)$,
for diagonal splitting preserve helicity in the $z\to 1$ limit.
Therefore, the collinear singularities have the form
 \begin{equation}
\label{Colsing}
{\cal A}^{\rm loop}_{\rm col}=
-\left(g\over 4\pi\right)^2
\sum_a^n {\gamma {(a)}\over \epsilon}
  {\cal A}^{\rm tree}.
\end{equation}
The constant $\gamma(a)$ represents the contribution from virtual graphs
to the Altarelli-Parisi kernel $P_{a/a}(\xi)$ or equivalently
it is defined by the
behavior of $P_{a/a}(\xi)$ near $\xi = 1$
\begin{equation}
\int_z^1 d\xi\ P_{a/a}(\xi)
= 2 C(a) \ln(1-z) + \gamma(a) + {\cal O}\left(1-z\right),
\label{gammadef}
\end{equation}
where $C(a)$ is the color charge of parton $a$.
Specifically, $C(a)$ and $\gamma(a)$ are
\begin{equation}
\begin{array}{lll}
C(g) =  N_c\, , &
\gamma(g) = {11N_c - 2 N_{\rm f} \over 6},& {\rm for\ \  gluons},\\
C(q) = { N_c^2-1 \over 2 N_c},&
\gamma(q) = {3 (N_c^2-1) \over 4 N}, & {\rm for\ \  quarks}.
\end{array}
\end{equation}

  The structure of multiple soft
emission from hard processes in QED was investigated by Gammer and Yennie
\cite{Gam73}. They have shown that the energetic electrons
participating in a hard process receive an eikonal phase
factor. In quantum chromodynamics, the situation is very similar
except that the eikonal factor is a matrix  equal
to the path order product of the matrix-valued gluon field
\cite{ColSop81,Cia81,March88}.

For one soft gluon, the main result is very simple:
it states that the singular contributions
come from a  configuration where the soft gluons are attached
to external legs of the graphs. Therefore,
the soft contribution can easily be calculated in terms
of the Born amplitude.
 For example, let us consider the
 contribution of a virtual soft gluon to the matrix element
of a process with $n$ external legs, $2\to (n-2)$.

The insertion of a soft gluon which connects the external legs $i$ and
$j$ has a twofold effect:

\noindent
First, the  Born amplitude gets rotated in the
color space by
 the insertion
of the color matrices appearing in the two vertices of the
soft line connecting
 the hard lines  $i$ and $j$
\begin{equation}
{\cal A}(2 \to n-2)_{c_1 c_2..c_n} \to
\sum_{a, c_i', c_j'}
t^{a}_{ c_i  c_i'}
t^{a}_{ c_j  c_j'}
{\cal A}(2 \to n-2)_{c_1.. c_i'..c_j'..c_n}
\end{equation}
where the labels of the initial partons are  $(n-1)$ and $n$;
 $c_i$'s denote color indices for the external partons
 and
$t^{a}_{c_i c_i'}$ is the SU(3) generator matrix for the color
representation of line $i$.  That is, $t^{a}_{ij}$ is
$(1/2) \lambda_{ij}^a$        for an outgoing quark,
$-(1/2) \lambda_{ij}^{*\, a}$ for an outgoing antiquark, and
$i f_{aij}$                   for an outgoing  gluon.
For an incoming parton, one can use the same formula as long
as one uses the conjugate color representations,
$-(1/2) \lambda_{ij}^{*\, a}$ for a quark,
$(1/2) \lambda_{ij}^a$        for an antiquark, and
$i f_{aij}$                   for a gluon.

Secondly, after carrying  out the loop integral and dropping singular
terms corresponding to collinear configurations, we  pick up
 the same eikonal factor as in QED, i.e.

\begin{equation}
- \left( {g\over 4\pi}\right)^2 c_{\Gamma}
\frac{1}{\epsilon^2}\left(-{ \mu^2\over s_{ij}  }\right)^{\epsilon},
\label{softsin}
\end{equation}
where we introduced the short hand notation
\begin{equation}
c_\Gamma={(4\pi)^{\epsilon}}
\frac{\Gamma^2(1-\epsilon)\Gamma(1+\epsilon)}
{\Gamma(1-2\epsilon)}\ .
\end{equation}
 Using unitarity this form of the soft factor can be confirmed
 by integrating the gluon momenta over  the Bremsstrahlung
eikonal factor \cite{Kun92}.

To obtain the complete
soft singular contributions, one should consider all
configurations
where a  gluon connects two  external legs.
 So   the soft singular terms of  loop helicity
amplitudes can be written as
\begin{equation}\label{softloop}
{\cal A}^{\rm loop}_{\rm soft}(2 \to n-2)_{c_1 c_2..c_n} =
- \sum_{i<j}
\left( {g\over 4\pi}\right)^2 c_{\Gamma}\frac{1}
{\epsilon^2}\left(-{ \mu^2\over
s_{ij}  }\right)^{\epsilon}
\sum_{a, c_i', c_j'}
t^{a}_{ c_i  c_i'}
t^{a}_{ c_j  c_j'}
{\cal A}^{\rm tree}(2 \to n-2)_{c_1.. c_i'..c_j'..c_n}
\end{equation}
The color matrix
of the Born term is an invariant tensor, therefore the sum of
insertions  to all
external lines must vanish  (soft identity \cite{Kun92}).
The soft structure given by (\ref{softloop}) can easily be decomposed
into  color subamplitudes.
As one possible simple test of the correctness of this algorithm, one
can use it to reproduce the
 singular terms of the one loop
helicity amplitudes of all $2\to 2$ subprocesses \cite{ZKASZT93}.

\def\d{{\rm d}}
\def\e{{\rm e}}
\def\i{{\rm i}}
\def\Am{{\cal A}}
\def\F{{\cal F}}
\def\I{{\cal I}}
\def\M{{\cal M}}
\def\S{{\cal S}}
\def\Li2{{\rm Li}_2}
\def\phys{{\rm phys}}
\def\sing{{\rm sing}}
\def\fin{{\rm fin}}
\def\ds{\displaystyle}
\def\to{\rightarrow}
\def\ms{$\overline{{\rm MS}}$}
\def\lQCD{$\Lambda_{{\rm QCD}}$}
\def\l{\langle}
\def\r{\rangle}
\def\vspaceinarray{\nonumber ~&~&~\\}
\def\nn{\nonumber}
\def\limes#1{\mathrel{\mathop{\lim}\limits_{#1}}}
\def\A#1#2{\l#1#2\r}
\def\B#1#2{[#1#2]}
\def\s#1#2{s_{#1#2}}
\def\mus#1#2{\left(-\frac{\mu^2}{s_{#1#2}}\right)^\varepsilon}
\def\N{N_c}
\def\Nf{N_{\rm f}}
\def\qb{\bar{q}}
\def\Qb{\bar{Q}}

\newcommand{\bq}{{\bar q}}
\newcommand{\bQ}{{\bar Q}}
\newcommand{\ba}{{\bar a}}
\newcommand{\la}{\langle}
\newcommand{\ra}{\rangle}
\newcommand{\psisi}{\psi^{\rm si}}

\section{Singular terms of the one loop color  subamplitudes}
\setcounter{equation}{0}

In this  section we give
the singular soft and collinear terms of the one loop color subamplitudes
of the $2\to 3$ subprocesses $qQ\to qQg$ and $q\overline{q}\to ggg$.
One can obtain the result for  the $qq\to qqg$  equal flavor
subprocess from the unequal flavor process
with simple permutations.
We do not give the result for the subprocess
$gg\to ggg$ since it was  obtained earlier  in
\cite{Ber93} and we found agreement. We note that  we calculated the
one loop helicity amplitudes of the subprocess $qQ\to qQg$
with standard diagrammatic method
\cite{ZKASZT94} and could confirm the correctness
of the Born level algorithm for the singular terms also in this case.

\subsection{Subprocess class $0 \to \qb \Qb Q q g$ }

\noindent
Here we present the singular terms of  the one loop
virtual corrections of the helicity color subamplitudes of the process
$0 \to \qb \Qb Q q g$.
The momenta of the partons are labeled as
\begin{equation}
0 \to {\rm antiquark_1}(\overline{q}) + {\rm antiquark_2}(\overline{Q})
+ {\rm quark_2}(Q) + {\rm quark_1}(q) + {\rm gluon}(g)\ .
\label{label4q1g}
\end{equation}
The color structure of the amplitudes is the same at one loop as at tree
level:
\begin{eqnarray}
\label{treeA}
\Am^{\rm tree}(\qb,\Qb;Q,q,g)&=&
g^3\left[\sum_{(q_1\ne q_2)\in \{q,Q\}}
(T^{g})_{q_1\qb_2}\delta_{q_2\qb_1}
a^{(0)}_{q_1\qb_2}(h_{\bq},h_{\bQ};h_Q,h_q,h_5)\right. \\ \nn
&& \quad-\left.\sum_{(q_1\ne q_2)\in \{q,Q\}}
\frac{1}{N_c}(T^{g})_{q_1\qb_1}\delta_{q_2\qb_2}
a^{(0)}_{q_1\qb_1}(h_{\bq},h_{\bQ};h_Q,h_q,h_5)\right] \\
\vspaceinarray
\label{oneloopA}
\Am^{\rm loop}(\qb,\Qb;Q,q,g)&=&
g^3\left({g\over 4\pi}\right)^2\left[\sum_{(q_1\ne q_2)\in \{q,Q\}}
(T^{g})_{q_1\qb_2}\delta_{q_2\qb_1}
a^{(1)}_{q_1\qb_2}(h_{\bq},h_{\bQ};h_Q,h_q,h_5)\right. \\ \nn
&& \quad-\left.\sum_{(q_1\ne q_2)\in \{q,Q\}}
\frac{1}{N_c}(T^{g})_{q_1\qb_1}\delta_{q_2\qb_2}
a^{(1)}_{q_1\qb_1}(h_{\bq},h_{\bQ};h_Q,h_q,h_5)\right] \\
\end{eqnarray}
The tree-level $a^{(0)}_{ij}$ subamplitudes are:
\begin{eqnarray}
a^{(0)}_{ij}(h_{\bq},h_{\bQ};h_Q,h_q,+) &=&
p_a(h_{\bq},h_{\bQ};h_Q,h_q,+)\,\frac{\A ij}{\A ig \A gj}, \\
a^{(0)}_{ij}(h_{\bq},h_{\bQ};h_Q,h_q,-) &=&
p_a(h_{\bq},h_{\bQ};h_Q,h_q,-)\,\frac{\B ij}{\B ig \B gj}.
\end{eqnarray}
The helicity dependence is completely absorbed in the factor $ p $
which is given by
\begin{eqnarray}
p_a(h_{\bq},h_{\bQ};h_Q,h_q,+) &=& \mp\i\frac{\A mn^2}{\A q\qb \A Q\Qb},
\label{pa1} \\
p_a(h_{\bq},h_{\bQ};h_Q,h_q,-) &=& \pm\i\frac{\B mn^2}{\B q\qb \B Q\Qb},
\label{pa2}
\end{eqnarray}
where $m$ and $n$ labels those (anti)quarks which have opposite
helicity to the gluon and the upper signs apply when $m$ and $n$ are
both quarks or antiquarks, while the lower signs apply when one of
them is a
quark and the other is an antiquark.

Using these expressions, the singular parts of the renormalized one-loop
subamplitudes can be written in a simple form:
\begin{eqnarray}
a^{(1), {\rm sing}}_{ Q \bq} &=&
-\frac{c_\Gamma}{\varepsilon^2} \left\{
\N a^{(0)}_{ Q \bq}
\left[\mus \bq g + \mus \bQ q + \mus Q g \right] \right. \\ \nn
&& \qquad -\frac{1}{\N} a^{(0)}_{ Q \bq}
\left[-\mus \bq \bQ + \mus \bq Q + \mus \bq q \right. \\ \nn
&& \qquad \qquad \qquad \left. + \mus \bQ Q + \mus \bQ q - \mus Q q
\right]
\\ \nn
&& \qquad -\frac{1}{\N} a^{(0)}_{ Q \bQ}
\left[-\mus \bq \bQ + \mus \bq g + \mus \bQ q - \mus q g \right] \\ \nn
&& \qquad -\frac{1}{\N} a^{(0)}_{ q \bq}
\left.\left[\mus \bQ q - \mus \bQ g - \mus Q q + \mus Q g \right]
\right\}
\\ \nn
&&-\frac{1}{\varepsilon}
\left( 4 \gamma(q) + \gamma(g) \right)
a^{(0)}_{ Q \bq}
\end{eqnarray}

and

\begin{eqnarray}
a^{(1), {\rm sing}}_{ Q \bQ} &=&
-\frac{c_\Gamma}{\varepsilon^2} \left\{
\N a^{(0)}_{ Q \bQ}
\left[\mus \bq q + \mus \bQ g + \mus Q g \right] \right. \\ \nn
&& \qquad -\frac{1}{\N} a^{(0)}_{ Q \bQ}
\left[-\mus \bq \bQ + \mus \bq Q + \mus \bq q \right. \\ \nn
&& \qquad \qquad \qquad \left. + \mus \bQ Q + \mus \bQ q - \mus Q q
\right]
 \\ \nn
&& \qquad +\N a^{(0)}_{ Q \bq}
\left[\mus \bq \bQ - \mus \bq q - \mus \bQ g + \mus q g \right] \\ \nn
&& \qquad +\N a^{(0)}_{ q \bQ}
\left.\left[-\mus \bq q + \mus \bq g + \mus Q q - \mus Q g \right]
\right\}
\\ \nn
&&-\frac{1}{\varepsilon} \left( 4 \gamma(q) + \gamma(g) \right)
a^{(0)}_{ Q \bQ}. \\
\vspaceinarray \nn
\end{eqnarray}
The remaining subamplitudes can be obtained by simple
label permutations

\begin{equation}
a^{(1), {\rm sing}}_{ q \bQ} = a^{(1), {\rm sing}}_{ Q \bq}
\lower 4pt\hbox{$|_{ \bq\leftrightarrow  \bQ,  Q\leftrightarrow  q}$} \qquad
a^{(1), {\rm sing}}_{ q \bq} = a^{(1), {\rm sing}}_{ Q \bQ}
\lower 4pt\hbox{$|_{ \bq\leftrightarrow  \bQ,  Q\leftrightarrow  q}$}
\end{equation}

\subsection{Subprocess class  $ 0 \to q ggg \qb $ }

\noindent
In this section we present the singular terms of  the one-loop
virtual corrections of the helicity color
 subamplitudes of the process  $ 0 \to q ggg \qb $

\noindent
We use the the momentum labels as follows
\begin{equation}
0 \to {\rm quark}(q) + {\rm gluon}(1)
+ {\rm gluon}(2) + {\rm gluon}(3) + {\rm antiquark}(\overline{q})\ .
\label{label2q3g}
\end{equation}
The color structure of the amplitudes at tree level is given by
 \begin{equation}
\label{treeB}
\Am^{\rm tree}(q,g,g,g,\qb)=
g^3\sum_{{\rm perm~}(1,2,3)}
(T^{g_1}T^{g_2}T^{g_3})_{q\qb} b^{(0)}_{123}(h_q,h_1,h_2,h_3,h_{\bq}),
\end{equation}
while at one loop one finds
\begin{eqnarray}
\label{oneloopB}
\Am^{\rm loop}(q,g,g,g,\qb)&=&
g^3\left({g\over 4\pi}\right)^2\left[\sum_{{\rm perm~}(1,2,3)}\right.
(T^{g_1}T^{g_2}T^{g_3})_{q\qb} b^{(1)}_{123}(h_q,h_1,h_2,h_3,h_{\bq}) \\
\nn &&
\quad ~~~~~~~ +\sum_{i=1,2,3} (T^{g_i})_{q\qb}\delta_{g_kg_l}
b^{(1)}_i(h_q,h_1,h_2,h_3,h_{\bq}) \\ \nn
&& \quad ~~~~~~~
+\delta_{q\qb}\tr(T^{g_1}T^{g_2}T^{g_3})b^{(1)}_q(h_q,h_1,h_2,h_3,h_{\bq})
\\ \nn
&& \quad ~~~~~~~ \left.
+\delta_{q\qb}\tr(T^{g_3}T^{g_2}T^{g_1})b^{(1)}_\bq(h_q,h_1,h_2,h_3,h_{\bq})
\right].
\end{eqnarray}
In this equation $k$ and $l$ are the two indices from the set
$\{1,2,3\}\setminus i$ for a given $i\in \{1,2,3\}$.

The tree-level $b^{(0)}_{123}$ subamplitudes vanish for helicity
configurations $(\pm,h_1,h_2,h_3,\pm)$ and $(\mp,h,h,h,\pm)$.
When there is one gluon with negative helicity (denoted by $I$) and
two with positive helicity, then
\begin{eqnarray}
b^{(0)}_{123}(-,h_1,h_2,h_3,+) &=&
\frac{p_b(-,h_1,h_2,h_3,+)}{\A  q 1 \A 12 \A 23 \A 3 \bq  \A  \bq  q },
\label{res:b0(m123p)} \\
b^{(0)}_{123}(+,h_1,h_2,h_3,-) &=&
\frac{p_b(+,h_1,h_2,h_3,-)}{\A  q 1 \A 12 \A 23 \A 3 \bq  \A  \bq  q },
\label{res:b0(p123m)}
\end{eqnarray}
where
\begin{eqnarray}
p_b(-,h_1,h_2,h_3,+) &=& \i \A  q I^3 \A  \bq I  , \label{pb1}\\
p_b(+,h_1,h_2,h_3,-) &=& -\i \A  q I \A  \bq I^3. \label{pb2}
\end{eqnarray}
The subamplitudes for processes with opposite helicities can be
obtained from (\ref{res:b0(m123p)} - \ref{pb2}) by replacing $\A ~~$
with  $\B ~~$.

Using these expressions, the singular parts of the renormalized one-loop
subamplitudes can again be written in a simple form. For the $b^{(1),
{\rm sing}}_{ijk}$ amplitudes one finds
\begin{eqnarray}
b^{(1), {\rm sing}}_{123} &=& -\frac{c_\Gamma}{\varepsilon^2}
\left\{\N\left[\mus q 1+\mus12+\mus23+\mus 3 \bq \right]\right. \\ \nn
&&\left. \qquad -\frac{1}{\N}\mus q \bq \right\} b^{(0)}_{123},\\ \nn
&& -\frac{1}{\varepsilon}
\left\{2 \gamma(q) + 3 \gamma(g) \right\} b^{(0)}_{123}.
\end{eqnarray}
and the other permutations are obtained by simple rearrangement of
indices:
\begin{equation}
b^{(1), {\rm sing}}_{132} = b^{(1), {\rm sing}}_{123}
\lower 4pt\hbox{$|_{2\leftrightarrow 3}$}, \qquad
b^{(1), {\rm sing}}_{231} = b^{(1), {\rm sing}}_{123}
\lower 4pt\hbox{$|_{1\to 2, 2\to 3, 3\to 1}$},
\end{equation}
\begin{equation}
b^{(1), {\rm sing}}_{213} = b^{(1), {\rm sing}}_{123}
\lower 4pt\hbox{$|_{1\leftrightarrow 2}$}, \qquad
b^{(1), {\rm sing}}_{312} = b^{(1), {\rm sing}}_{123}
\lower 4pt\hbox{$|_{1\to 3, 2\to 1, 3\to 2}$}, \qquad
b^{(1), {\rm sing}}_{321} = b^{(1), {\rm sing}}_{123}
\lower 4pt\hbox{$|_{1\leftrightarrow 3}$}.
\end{equation}
For the $b^{(1), {\rm sing}}_i$ amplitudes one finds
\begin{eqnarray}
b^{(1), {\rm sing}}_1 &=& -\frac{c_\Gamma}{\varepsilon^2}
\left\{\left[-\mus13+\mus1 \bq +\mus23-\mus2 \bq \right]b^{(0)}_{123}
\right. \\
\vspaceinarray \nn && \qquad \qquad \left.
+[2\leftrightarrow 3] b^{(0)}_{132}
+[ q \leftrightarrow  \bq ] b^{(0)}_{321}
+[2\leftrightarrow 3,  q \leftrightarrow  \bq ] b^{(0)}_{231}\right\},
\end{eqnarray}
\begin{eqnarray}
b^{(1), {\rm sing}}_2 &=& -\frac{c_\Gamma}{\varepsilon^2}
\left\{\left[\mus q 2-\mus q 3-\mus12+\mus13\right]b^{(0)}_{132}\right. \\
\vspaceinarray \nn && \qquad \qquad \left.
+[1\leftrightarrow 3] b^{(0)}_{312}
+[ q \leftrightarrow  \bq ] b^{(0)}_{231}
+[1\leftrightarrow 3,  q \leftrightarrow  \bq ] b^{(0)}_{213}\right\},
\end{eqnarray}
\begin{eqnarray}
b^{(1), {\rm sing}}_3 &=& -\frac{c_\Gamma}{\varepsilon^2}
\left\{\left[-\mus q 2+\mus q 3+\mus12-\mus13\right]b^{(0)}_{123}\right. \\
\vspaceinarray \nn && \qquad \qquad \left.
+[1\leftrightarrow 2] b^{(0)}_{213}
+[ q \leftrightarrow  \bq ] b^{(0)}_{321}
+[1\leftrightarrow 2,  q \leftrightarrow  \bq ] b^{(0)}_{312}\right\},
\end{eqnarray}
\begin{eqnarray}
b^{(1), {\rm sing}}_ q  &=& -\frac{c_\Gamma}{\varepsilon^2}
\left\{\left[-\mus q 3+\mus q  \bq +\mus13-\mus1 \bq
\right]b^{(0)}_{123} \right. \\
\vspaceinarray \nn && \qquad \qquad \left.
+(1\to 2, 2\to 3, 3\to 1) + (1\to 3, 2\to 1, 3\to 2)\right\},
\end{eqnarray}
\begin{eqnarray}
b^{(1), {\rm sing}}_ \bq  &=& -\frac{c_\Gamma}{\varepsilon^2}
\left\{\left[-\mus q 2+\mus q  \bq +\mus12-\mus1 \bq
\right]b^{(0)}_{132} \right. \\
\vspaceinarray \nn && \qquad \qquad \left.
+(1\to 2, 2\to 3, 3\to 1) + (1\to 3, 2\to 1, 3\to 2)\right\},
\end{eqnarray}

\section{Soft limits of the cross sections of $2\to 4$ scattering
processes}
\setcounter{equation}{0}

The soft singular terms in the one loop radiative corrections
of the $2\to 3$ hard processes determine the soft momentum dependence
of a soft gluon emission from the hard partons that is
the soft gluon momentum limit of  general $2\to 4$ processes

\begin{equation}
{\rm parton}(p_1)+{\rm parton}(p_2) \to {\rm parton}(p_3)
 + {\rm parton}(p_4)
 + {\rm parton}(p_5) + {\rm gluon}(k).
\end{equation}
 The soft momentum dependence factorizes into the
eikonal factor
\begin{equation}
e(m,n) =  g^2{p_n \cdot p_m \over p_n \cdot k \, p_m\cdot k}
\end{equation}
so that  we may write in the soft limit \cite{Kun92,Cia81}
\begin{equation}
\sum_{\rm color}|M(p_1,p_2,p_3,p_4,p_5,k)|^2 {\to}
g^4 \sum_{\rm n<m} e(m,n) \psi_{\rm mn}(p_i,h_i).
\end{equation}
Integrating the {\it d}-dimensional phase space over the
gluon momenta we obtain the same singular factor as given
by eq. (\ref{softsin}). Since the soft singularities cancel
we can obtain the spin dependent
 $\psi_{\rm nm} (p_i,h_i)$
 coefficient functions of the eikonal
factors
by calculating the singular contributions of the
virtual corrections to the colored summed
cross section of the $2\to 3$ hard scattering process.

On the other hand, the calculation of the spin dependent
$\psi_{\rm mn}(p_i,h_i)$ functions can also
be carried out
directly from the soft Bremsstrahlung equation (see ref.\cite{Kun92}).
We carried out the calculation using both methods, in order to test the
correctness of our results.
It is an interesting feature of the soft limit that
 the spin dependence of the  $\psi_{\rm mn}$ functions is factorizable:
it can be
absorbed into  a  simple factor  $ p $ (see eqs. (\ref{pa1},\ref{pa2},
\ref{pb1},\ref{pb2}) )  which is
independent of the color structure.
  The helicity dependence of
 the $\psi_{\rm mn}(p_i,h_i)$ functions can be given as
\begin{equation}
\psi_{\rm mn} (p_i,h_i) =  |p (p_i,h_i)|^2 \psisi_{\rm mn}(p_i),
\end{equation}
where $\psisi$ is the spin independent part of $\psi_{\rm
mn}(p_i,h_i)$.

The angular ordering approximation to the soft gluon distribution
in the hard process is derived by the azimuthal averaging over the
eikonal factor. Shower Monte Carlo programs use
this piece of information, therefore, it is of interest to know
the soft coefficient functions. In particular one can develop
an initial condition for the Monte Carlo programs where
for a given $2\to 3$ hard parton process a cone for the
first gluon emission from each external line is chosen according
to the probabilities defined with the help of soft emission.
Therefore to obtain analytic expressions for the
$\psi_{nm}$ functions is of phenomenological interest.
We note that 2-jet and 3-jet initial conditions are worked out
for $e^+e^-$ shower Monte Carlo programs but the 3-jet initial conditions
are not yet developed for hadron-hadron and electron-proton
collisions.

\subsection{ Subprocess class  $ 0 \rightarrow \bq \bQ Q q g (g) $}

\noindent
The momenta of the partons are labeled as in (\ref{label4q1g})
Then for the spin independent part of the soft coefficients we  obtain
the  expressions

\begin{eqnarray}
\psisi_{\bq \bQ}({h_i}) & = &
\frac{V}{N_c^2} \left( s_{qg}s_{Qg}s_{\bq g}s_{\bQ g} \right) ^{-1} \times \\
& & \left[ ( N_c^2+3) s_{\bq g}s_{\bQ g}s_{qQ} -
          2 s_{\bq g}s_{qg}s_{Q\bQ} - 2 s_{q\bq}s_{\bQ g}s_{Qg} \right. \nn \\
& & + \left. (N_c^2-3) \left( s_{\bq g}s_{q\bQ}s_{Qg} +
      s_{\bq Q}s_{\bQ g}s_{qg} - s_{\bq \bQ}s_{qg}s_{Qg} \right)
\right] \nn  \\
\psisi_{q \bq}({h_i})  & = &
-\frac{V}{N_c^2} \left( s_{qg}s_{Qg}s_{\bq g}s_{\bQ g} \right) ^{-1}
\times   \\
& & \left[ 2 s_{\bq \bQ}s_{qg}s_{Qg} + 2 s_{qQ}s_{\bQ g}s_{\bq g} -
         (N_c^2 +1) s_{\bQ g}s_{q\bq}s_{Qg} \right.  \nn \\
& & - \left. s_{Q\bQ}s_{\bq g}s_{qg} +
        (N_c^2-2) \left( s_{\bq Q}s_{\bQ g}s_{qg} +
        s_{\bq g}s_{q\bQ}s_{Qg} \right) \right] \nn \\
\psisi_{\bq Q}({h_i})  & = &
\frac{V}{ N_c^2} \left( s_{qg}s_{Qg}s_{\bq g}s_{\bQ g} \right) ^{-1}\times \\
& & \left[ (N_c^2-3) \left( s_{\bq \bQ}s_{qg}s_{Qg} +
       s_{\bq g}s_{\bQ g}s_{qQ} - s_{\bq Q}s_{\bQ g}s_{qg} \right)
\right.  \nn \\
&  & + (2 -N_c^2) \left( s_{q\bq}s_{Qg}s_{\bQ g} +
        s_{Q\bQ}s_{qg}s_{\bq g} \right)  \nn \\
& & + \left. (N_c^4 - 3 N_c^2 +3) s_{\bq g}s_{q\bQ}s_{Qg} \right] \nn \\
\psisi_{\bq g}({h_i})  & = &
V \left( s_{qg}s_{Qg}s_{\bq g}s_{\bQ g} \right) ^{-1}  \times \\
& & \left[ (N_c^2-2) s_{\bq Q}s_{\bQ g}s_{qg} - s_{\bQ g}s_{q\bq}s_{Qg}
         + 2 s_{\bq \bQ}s_{qg}s_{Qg} \right]. \nn
\end{eqnarray}
Out of these four $\psisi_{mn}$ functions one can  obtain all the ten
functions by  changing the labels correspondingly
\begin{itemize}
\item  $ \psisi_{qQ} =
 \psisi_{\bq \bQ}$ with $ q \leftrightarrow \bq, Q \leftrightarrow \bQ $
\item $ \psisi_{q \bQ} =
 \psisi_{\bq Q} $ with $ q \leftrightarrow Q, \bq \leftrightarrow \bQ $
\item $ \psisi_{qg} =
 \psisi_{\bq g}$ with $ q \leftrightarrow \bq, Q \leftrightarrow \bQ $
\item $ \psisi_{Q \bQ} =
 \psisi_{q \bq} $ with $ q \leftrightarrow Q, \bq \leftrightarrow \bQ $
\item $\psisi_{Qg} =
 \psisi_{\bq g} $ with $ \bq \leftrightarrow Q, \bQ \leftrightarrow q $
\item $ \psisi_{\bQ g} =
 \psisi_{\bq g} $ with $ q \leftrightarrow Q, \bq \leftrightarrow \bQ $
\end{itemize}
The spin summed coefficient functions are
\begin{equation}
\psi_{mn} = \frac{2}{s_{Q\bQ} s_{q\bq} } \left( s_{Qq}^2 + s_{\bQ q}^2
+ s_{Q \bq}^2 + s_{\bQ \bq}^2 \right) \psisi_{mn}
\end{equation}

\subsection{ Subprocess class  $ 0 \rightarrow q {\bar q} q {\bar q} g (g) $}

\noindent
In the case of equal flavor,
we should use the same amplitudes as in the case of unequal
flavor but we should antisymmetrize it in the momenta
$\overline{Q}\leftrightarrow
\overline{q} $ and in the corresponding color and helicity labels.

The $\psi_{mn}(h_i)$ functions then can be written in the  form

\begin{eqnarray}
\psi_{\rm mn}(h_i) &=&
- \frac{V}{N} \left( |p_a(\bq,\bQ;Q,q;g)|^2 \ A_{\rm mn} +
|p_a(\bQ,\bq;Q,q;g)|^2 \ B_{\rm mn} \right. \nonumber \\
& & \left. \quad
- 2 {\rm Re} (p_a(\bq,\bQ;Q,q;g) p_a(\bQ,\bq;Q,q;g)^* C_{\rm mn}) \right) .
\label{ABC}
\end{eqnarray}
 To simplify the final expressions we use  the notation
  $$ \{ ij \} \equiv \frac{2 s_{ij}}{s_{ig}s_{jg}}, ~~~~~ f_{ij}
\equiv \frac{ \A i j}{\A i g \A g j} $$
but here
 $g$ denotes the hard gluon momentum, and $i,j$ can run over values
$q, Q,  \bq $ or $\bQ$.
Then the auxiliary functions $A,B,C$ appearing in (\ref{ABC})
can be written in the
simple form
\begin{eqnarray*}
A_{\bq\bQ} & = &
\left( \frac{3}{2 \N}-\frac{\N}{2} \right) \left( \{ q\bQ \} +
\{ \bq Q\} - \{ \bq\bQ \} \right) + \frac{1}{\N} \{ q \bq \}   \\
&-& \left( \frac{3}{2 \N} + \frac{\N}{2} \right) \{q Q \}
+ \frac{1}{\N} \{Q\bQ \} \\
B_{\bq\bQ} &=&
A_{\bq \bQ}\lower 4pt\hbox{$|_{\bq\leftrightarrow \bQ}$} \\
C_{\bq\bQ} & = &
\frac{V}{\N^2} \left( f_{Q \bQ} f^*_{Q \bq}
    + f_{q \bq} f^*_{q \bQ} \right)
    + V \left( f_{Q \bq} f^*_{Q \bQ}
    + f_{q \bQ} f^*_{q \bq} \right) \\
    &-& \frac{1}{\N^2} \left( f_{q \bq} f^*_{Q \bq}
    + f_{Q \bQ} f^*_{q \bQ} \right) \\
    &-&  \left( f_{q \bQ} f^*_{Q \bq}
    + f_{q \bq} f^*_{Q \bQ}
    + f_{q \bQ} f^*_{Q \bQ}
    + f_{Q \bq} f^*_{q \bq}
    + f_{Q \bQ} f^*_{q \bq}
    + f_{Q \bq} f^*_{q \bQ} \right) \\
A_{q \bq} &=&
\left( \frac{\N}{2} - \frac{1}{\N} \right)
\left( \{\bq Q \} + \{q \bQ \} \right) - \left( \frac{1}{2 \N} +
\frac{\N}{2} \right) \{q \bq \} \\
& + & \frac{1}{\N} \{ \bq \bQ \} + \frac{1}{\N} \{ q Q \} -
\frac{1}{2 \N} \{ Q \bQ \} \\
B_{q \bq} &=& A_{\bq Q} \lower 4pt\hbox{$|_{q\leftrightarrow Q}$} \\
C_{q \bq} &=&
- \frac{V}{\N^2} \left( f_{Q \bQ} f^*_{Q \bq}
      + f_{Q \bQ} f^*_{q \bQ} \right)
      + \frac{1}{\N^2} \left( f_{q \bq} f^*_{Q \bq}
      + f_{q \bq} f^*_{q \bQ}\right) \\
      &-& f_{q \bq} f^*_{q \bq}
      + V f_{Q \bQ} f^*_{Q \bQ}
      - \N^2 \left(f_{q \bQ} f^*_{Q \bQ}
      + f_{Q \bq} f^*_{Q \bQ} \right) \\
      &+&  \left( f_{q \bQ} f^*_{Q \bq}
      + f_{Q \bq} f^*_{Q \bQ}
      + f_{q \bQ} f^*_{Q \bQ}
      + f_{Q \bq} f^*_{q \bq}
      + f_{q \bQ} f^*_{q \bq}
      + f_{Q \bq} f^*_{q \bQ}\right) \\
A_{\bq Q} &=& \left( \frac{3}{2 \N} - \frac{\N}{2} \right)
\left( \{\bq \bQ \} + \{ q Q\} - \{\bq Q \} \right) \\
&+& \left( \frac{-3}{2 \N} + \frac{3 \N }{2} -
\frac{\N^3}{2} \right) \{q \bQ \} + \left( \frac{\N}{2} - \frac{1}{\N}
\right) \left( \{q \bq \} + \{Q \bQ \} \right) \\
B_{\bq Q} &=& A_{q \bq} \lower 4pt\hbox{$|_{q\leftrightarrow Q}$} \\
C_{\bq Q} &=&
 - \frac{V}{\N^2} \left( f_{Q \bQ} f^*_{q \bQ}
      + f_{q \bq} f^*_{q \bQ}\right)
      + \frac{1}{\N^2} \left( f_{Q \bQ} f^*_{Q \bq}
      + f_{q \bq} f^*_{Q \bq}\right) \\
      &-& f_{Q \bq} f^*_{Q \bq}
      + V f_{q \bQ} f^*_{q \bQ}
      - \N^2 \left(f_{q \bQ} f^*_{q \bq}
      + f_{q \bQ} f^*_{Q \bQ} \right)\\
      &+&  \left(f_{Q \bq} f^*_{Q \bQ}
      + f_{q \bq} f^*_{Q \bQ}
      + f_{q \bQ} f^*_{Q \bQ}
      + f_{Q \bq} f^*_{q \bq}
      + f_{Q \bQ} f^*_{q \bq}
      + f_{q \bQ} f^*_{q \bq}\right) \\
A_{\bq g} &=&
\frac{\N}{2} \{q \bq \} - \N \{\bq \bQ \} + \left( \frac{-\N^3}{2}
+ \N \right) \{\bq Q \} \\
B_{\bq g} &=& A_{\bq g} \lower 4pt\hbox{$|_{q\leftrightarrow Q}$} \\
C_{\bq g} &=&
 \N^2 \left( f_{Q \bq} f^*_{Q \bq}
    + f_{q \bQ} f^*_{Q \bQ}
    + f_{q \bq} f^*_{q \bq}
    - f_{Q \bq} f^*_{Q \bQ}
    - f_{Q \bq} f^*_{q \bq}
    - f_{q \bQ} f^*_{q \bq} \right) \\
    &+&  \left( f_{Q \bQ} f^*_{q \bQ}
    - f_{q \bq} f^*_{q \bQ}
    - f_{Q \bQ} f^*_{Q \bq}
    - f_{q \bq} f^*_{Q \bq}\right)
\end{eqnarray*}
In order to get the expressions for the other $\psisi_{mn}$ functions,
we have to do the same  relabeling procedure as in the previous
process now with the $ A, B$ and $ C$ functions .
The spin summed $\psi_{\rm mn}$ functions can be obtained by
\begin{eqnarray}
|p_a(\bq,\bQ;Q,q;g)|^2 &\to& \frac{2}{s_{Q\bQ} s_{q\bq}}
(s_{qQ}^2 + s_{\bq \bQ}^2 + s_{q\bQ}^2 + s_{\bq Q}^2)  \\
|p_a(\bQ,\bq;Q,q;g)|^2 &\to& \frac{2}{s_{q\bQ} s_{\bq Q}}
(s_{qQ}^2 + s_{\bq \bQ}^2 + s_{q\bq}^2 + s_{Q\bQ}^2) \nn \\
\lefteqn{ {\rm Re} (p_a(\bq,\bQ;Q,q;g)p_a(\bQ,\bq;Q,q;g)^* C_{\rm
mn})} \hspace{3 cm} ~ \nn \\
&\to&
\frac{s_{\bq \bQ}^2 + s_{qQ}^2}{ s_{q\bq}s_{Q\bQ}s_{q\bQ}s_{\bq Q}}
( s_{q\bq} s_{Q\bQ} + s_{q\bQ} s_{\bq Q} - s_{qQ}s_{\bq \bQ} ) {\rm
Re}(C_{\rm mn})\nn
\end{eqnarray}

\subsection{ Subprocess class  $ 0 \rightarrow q {\bar q} g g g (g) $}

\noindent
We use the the momentum labels as in (\ref{label2q3g})
and we introduce the short hand notation

$$
{\cal D} \equiv
s_{12}s_{13}s_{23}s_{1q}s_{1\bq}s_{2q}s_{2\bq}s_{3q}s_{3\bq}s_{q\bq}.
$$
\samepage
Then for the soft color coefficient functions we obtain
\begin{eqnarray}
\psisi_{q 1} &= &
\frac{V}{N_c} \frac{s_{q\bq}}{s_{1q}s_{1\bq}s_{2q}s_{2\bq}s_{3q}s_{3\bq}} \\
& + &  (N_c^5 - N_c) \frac{1}{s_{23}s_{1q}s_{q\bq}}
\left(\frac{1}{s_{12}s_{3\bq}} + \frac{1}{s_{13}s_{2\bq}} \right) \nn \\
& - & \frac{N_c V}{{\cal D}} \left\{
 - 3 \cdot s_{23} s_{1q}^2  s_{2\bq} s_{3\bq} -
   s_{23}^2 s_{1q} s_{1\bq} s_{q\bq}
 + s_{23} s_{1\bq}^2 s_{2q} s_{3q}  \right. \nn \\
& & \quad + \left[ s_{13}s_{1q}s_{2q}s_{2\bq}s_{3\bq}  +
   s_{23}s_{1q}s_{1\bq}s_{2\bq}s_{3q}
    + s_{13}s_{23}s_{1q}s_{2\bq}s_{q\bq} \right. \nn \\
& &  \quad ~~~ + \left. \left.  s_{13}^2 s_{2\bq}s_{2q}s_{q\bq} +
   s_{13} s_{1\bq} s_{2q}^2 s_{3\bq}
  - s_{13}s_{1q}s_{3q}s_{2\bq}^2 + (2 \leftrightarrow 3) \right] \right\}.\nn
\end{eqnarray}
{}From the above expression we get
 five other $\psisi$  functions by  changing the labels  of the momenta in
the spin independent part
\begin{itemize}
\item $\psisi_{q 2} : ( q,\bq,1,2,3 ) \rightarrow (q , \bq, 2,3,1)$
\item $\psisi_{q 3} : ( q,\bq,1,2,3 ) \rightarrow (q, \bq ,3,1,2) $
\item $\psisi_{\bq 1} : ( q,\bq,1,2,3 ) \rightarrow (\bq, q, 1,2,3) $
\item $\psisi_{\bq 2} : ( q,\bq,1,2,3 ) \rightarrow (\bq , q,2,3,1) $
\item $ \psisi_{\bq 3} : ( q,\bq,1,2,3 ) \rightarrow (\bq, q, 3,1,2 )$
\end{itemize}
There are still two additional independent coefficient functions which
cannot be obtained by symmetry properties
\begin{eqnarray}
\psisi_{q \bq} & = &
\frac{(1-N_c^4)}{N_c^3}
      \frac{s_{q\bq}}{s_{1q}s_{1\bq}s_{2q}s_{2\bq}s_{3q}s_{3\bq}}  \nn \\
& + & \frac{N_c V}{{\cal D}} \sum_{Z(123)} \left[
  - 2 \cdot s_{12} s_{3q}^2 s_{1\bq}s_{2\bq}
  - 2 \cdot s_{12} s_{3\bq}^2s_{1q}s_{2q}  \right. \\
& & \ + \ \left.\left[ s_{12}s_{3q}s_{3\bq} + s_{13}s_{23}s_{q\bq} \right]
\left[ s_{1q}s_{2\bq} + s_{1\bq} s_{2q} \right] \right] \nn \\
& - & \frac{V}{{\cal D} \ N_c}
 \sum_{Z(123)}  \left[ s_{12}s_{13}s_{2q}s_{3\bq}s_{q\bq}
  - 2 \cdot s_{23}s_{1q}s_{1\bq}s_{2\bq}s_{3\bq}   \right. \nn \\
 & & \ - \ \left.  2 \cdot s_{1q}^2 s_{23}s_{2\bq}s_{3\bq}
  + s_{12}s_{13}s_{23}s_{1q}s_{1\bq} + ( q \leftrightarrow \bq )
\right], \nn
\end{eqnarray}
where $Z(123)$ denotes the cyclic permutations and
\begin{eqnarray}
\psisi_{12} &=&
\frac{N_c^3 V}{s_{q\bq}}
\left( \frac{1}{s_{12}s_{13}s_{2q}s_{3\bq}} +
       \frac{1}{s_{12}s_{23}s_{1q}s_{3\bq}} +
       ( q \leftrightarrow \bq ) \right)  \\
& - & \frac{N_c V}{{\cal D}} \left\{ - 2 \cdot s_{12}s_{13}s_{23}
       \left[ s_{1q}s_{1\bq} + s_{2q}s_{2\bq} \right] \right. \nn \\
& & \quad - 2 \cdot s_{12}^2 s_{3q}s_{3\bq}s_{q\bq} -
      s_{13}s_{23}s_{q\bq} \left[ s_{1q}s_{2\bq} + s_{1\bq}s_{2q}
\right] \nn \\
& & \quad + 2 \cdot  s_{23} \left[ s_{1q}+s_{1\bq} \right]
       \left[ s_{1\bq}s_{2q}s_{3q} + s_{1q}s_{2\bq}s_{3\bq} \right]
\nn \\
& & \quad + \left. 2 \cdot  s_{13} \left[ s_{2q} + s_{2\bq} \right]
   \left[ s_{1\bq}s_{2q}s_{3\bq} + s_{1q}s_{2\bq}s_{3q} \right] \right\}\nn
\nopagebreak[4]
\end{eqnarray}
Spin summation results in
\begin{equation}
\psi_{\rm mn} = 2 \sum_{i \in \{1,2,3\}}
\left( s_{iq}^3 s_{i \bq} + s_{iq} s_{i \bq}^3 \right) \psisi_{\rm mn}
\end{equation}

\pagebreak

\subsection{ Subprocess class  $ 0 \rightarrow  g g g g g (g) $}

\noindent
We introduce the simple notation for labeling the quarks and gluons
as
\begin{equation}
0 \to {\rm gluon}(1)+{\rm gluon}(2) +
{\rm gluon}(3)+{\rm gluon}(4)+{\rm gluon}(5)
\end{equation}
Due to Bose symmetry it is sufficient to calculate only one of the
ten $\psi_{mn}$ functions
\begin{eqnarray}
\psisi_{13}  & = & 2 N_c^4 V
\sum_{P(2,4,5)} \frac{1}{s_{13}s_{32}s_{24}s_{45}s_{51}}   \\
& - &  24  \ N_c^2 V \left(
\frac{1}{s_{12}s_{14}s_{15}s_{24}s_{25}}
 + \frac{1}{s_{14}s_{15}s_{23}s_{25}s_{34}}  \right. \nn \\
& & \quad  \quad ~~~ + ~ \frac{1}{s_{14}s_{15}s_{23}s_{24}s_{35}}
          - \frac{1}{s_{12}s_{14}s_{15}s_{34}s_{35}} \nn \\
& & \quad  \quad ~~~ - ~ \frac{1}{s_{12}s_{24}s_{25}s_{34}s_{35}}
          + \frac{1}{s_{12}s_{15}s_{23}s_{34}s_{45}} \nn \\
& & \quad  \quad ~~~ + ~ \frac{1}{s_{12}s_{14}s_{23}s_{35}s_{45}}
          - \frac{2 s_{13}}{s_{12}s_{14}s_{15}s_{23}s_{34}s_{35}} \nn \\
& & \quad  \quad ~~~ - ~\left.
 \frac{s_{24}}{s_{12}s_{14}s_{23}s_{25}s_{34}s_{45}}
          - \frac{s_{25}}{s_{12}s_{15}s_{23}s_{24}s_{35}s_{45}} \right).\nn
\end{eqnarray}
Although in the expression  for $ \psisi_{13} $
 the symmetry $ 1 \leftrightarrow 3 $ is not manifest, using momentum
conservation we can check that it has the symmetry $1\leftrightarrow 3$.
We get the  other $\psisi_{mn} $
functions now simply by the usual replacements of the labels
of the  momenta, i.e. $\psisi_{13} \rightarrow \psisi_{12} $ by $ 3
\leftrightarrow 2 $.

The spin dependent part of the born amplitude is
given by
$$
p_d  =  \i \A mn^4
$$
if $m$ and $n$ are the only negative helicity gluons and by
$$
p_d  =  \i \B mn^4
$$
if $m$ and $n$ are the only positive helicity gluons. So,
for the helicity summed $\psi_{mn} $ functions
we get
\begin{equation}
\psi_{\rm  mn} = 2 \sum_{i < j} s_{ij}^4 \psisi_{\rm  mn}.
\end{equation}

\section{Conclusions}

We presented an algorithm of general validity, which enables us to
calculate the singular terms of any one-loop helicity amplitude in
QCD.
The correctness of the algorithm was tested by explicitly
demonstrating the cancellation of the soft singularities between the
loop and Bremsstrahlung contributions for spin dependent $
2 \to 3, \ 2 \to 4 $ processes.
We also emphasized the universality of the collinear singularities (see
(\ref{Colsing})).
Furtheron, we obtained short formul\ae \,  for the soft limit of the spin
dependent cross sections of all $ 2 \to 4$ processes. These results
represent a necessary input for shower Monte Carlo programs for the
production of three well-separated jets.

\def\np#1#2#3  {{\it Nucl. Phys. }{\bf #1} (19#3) #2}
\def\nc#1#2#3  {{\it Nuovo. Cim. }{\bf #1} (19#3) #2}
\def\pl#1#2#3  {{\it Phys. Lett. }{\bf #1} (19#3) #2}
\def\pr#1#2#3  {{\it Phys. Rev. }{\bf #1} (19#3) #2}
\def\prl#1#2#3  {{\it Phys. Rev. Lett.}{\bf #1} (19#3) #2}
\def\prep#1#2#3{{\it Phys. Rep. }{\bf #1} (19#3) #2}

\end{document}